\documentclass[twocolumn,showpacs,preprintnumbers,amsmath,amssymb]{revtex4}
\usepackage{mathrsfs}
\usepackage{amssymb}
\usepackage{graphicx}
\usepackage{dcolumn}
\usepackage{bm}
\usepackage{textcomp}
\usepackage{amsmath}
\usepackage[titletoc]{appendix}

\newcommand\ket[1]{\ensuremath{|#1\rangle}}

\newcommand\oprod[2]{\ensuremath{|#1\rangle\langle#2|}}
\newcommand\mean[1]{\ensuremath{\langle #1\rangle}}

\newcounter{RomanNumber}

\begin{document}
\title{General sending-or-not-sending twin field protocol for quantum key distribution with asymmetric source parameters}
\author{Xiao-Long Hu$ ^{1}$, Cong Jiang$ ^{1}$, Zong-Wen Yu$ ^{2}$, and Xiang-Bin Wang$ ^{1,3,4,5\footnote{Email
Address: xbwang@mail.tsinghua.edu.cn}} $
}


\affiliation{ \centerline{$^{1}$State Key Laboratory of Low
Dimensional Quantum Physics, Department of Physics,} \centerline{Tsinghua University, Beijing 100084,
People¡¯s Republic of China}
\centerline{$^{2}$Data Communication Science and Technology Research Institute, Beijing 100191, China}
\centerline{$^{3}$ Synergetic Innovation Center of Quantum Information and Quantum Physics, University of Science and Technology of China}
\centerline{  Hefei, Anhui 230026, China
 }
\centerline{$^{4}$ Jinan Institute of Quantum technology, SAICT, Jinan 250101,
People¡¯s Republic of China}
\centerline{$^{5}$ Shenzhen Institute for Quantum Science and Engineering, and Physics Department,}
\centerline{ Southern University of Science and Technology, Shenzhen 518055, China}
}
\begin{abstract}
The sending-or-not-sending (SNS) protocol of the twin-field quantum key distribution (TFQKD) can tolerant large misalignment error and its key rate can exceed the linear bound of repeaterless QKD.
But the original SNS protocol requires the two users to use the same source parameters. Here we propose a general protocol with asymmetric source parameters and give the security proof of this protocol.
Our general protocol has a much better performance than that of the original SNS protocol when the channel is asymmetric.

\end{abstract}


\pacs{
03.67.Dd,
42.81.Gs,
03.67.Hk
}
\maketitle

\section{Introduction}
Quantum key distribution provides a method for unconditionally secure communication~\cite{bennett1984quantum,lo1999unconditional,shor2000simple,mayers2001unconditional,gisin2002quantum,gisin2007quantum,renner2008security,scarani2009security,koashi2009simple} between two parties, Alice and Bob.
Combined with the decoy-state method~\cite{inamori2007unconditional,gottesman2004security,hwang2003quantum,wang2005beating,lo2005decoy,adachi2007simple} and measurement-device-independent QKD (MDIQKD) protocol~\cite{lo2012measurement,braunstein2012side}, QKD can overcome the security loophole from the nonideal single-photon sources~\cite{huttner1995quantum,yuen1996quantum,brassard2000limitations} and imperfect detection devices~\cite{lydersen2010hacking,gerhardt2011full} and has developed rapidly both in theory~\cite{wang2007quantum,hayashi2007upper,wang2013three,sasaki2014practical,curty2014finite,xu2013practical,xu2014protocol,song2012finite,zhou2014tightened,yu2015statistical,zhou2016making,jiang2016measurement,jiang2017measurement,zhou2017obtaining,huang2018quantum,chau2018decoy,hu2018measurement,wang2018prefixed} and experiment~\cite{rosenberg2007long,schmitt2007experimental,peng2007experimental,boaron2018secure,yuan2007unconditionally,wang2008experimental,peev2009secoqc,dixon2010continuous,sasaki2011field,frohlich2013quantum,rubenok2013real,liu2013experimental,da2013proof,chan2014modeling,tang2014experimental,tang2014measurement,takesue2015experimental,wang2015phase,pirandola2015high,comandar2016quantum,wang2017measurement,liao2017satellite,liao2018satellite}.
The maximum distance of decoy-state MDIQKD has been experimentally increased to 404 kilometers~\cite{yin2016measurement}.
But the key rate of BB84, MDIQKD protocol, or any modified version of these protocols cannot exceed the linear bounds of repeaterless QKD, such as the TGW bound~\cite{takeoka2014fundamental} and the PLOB bound~\cite{pirandola2017fundamental}.

Recently, a new protocol named twin-field quantum key distribution (TFQKD) was proposed~\cite{lucamarini2018overcoming} whose key rate dependence on the channel transmittance $\eta$ is $R\sim O(\sqrt\eta)$.
Following this protocol, many variants of TFQKD were proposed~\cite{wang2018twin,tamaki2018information,ma2018phase,lin2018simple,cui2019twin,curty2018simple,lu2019twin,grasselli2019practical,xu2019general,zhang2019twin,zhou2019asymmetric,maeda2019repeaterless} and some experiments of TFQKD were demonstrated~\cite{minder2019experimental,liu2019experimental,wang2019beating,zhong2019proof}.
Among those protocols, one efficient protocol named the sending-or-not-sending (SNS)  protocol~\cite{wang2018twin} has the advantages of unconditionally security under coherent attacks and it can tolerant large misalignment error.
And the SNS protocol with finite data size has been studied~\cite{yu2019sending,jiang2019unconditional}.
However, the security proof of the SNS protocol requires the condition that the two users, Alice and Bob, use the same source parameters, such as the intensities of signal and decoy sources and the probability for sending coherent pulses in the $Z$ windows.

Here we propose a general SNS protocol where Alice and Bob are not required to use the same source parameters.
We give a security proof for this general protocol. Then we apply our general protocol to the case of asymmetric channels, i.e., the channel between Alice and Charlie (we will call it ``Alice's channel'' for simplicity in this paper) and that between Bob and Charlie (we will call it ``Bob's channel'' for simplicity in this paper) are not the same.
The numerical results show that in this case the key rate of our general protocol is much higher than that of the original SNS protocol.

This paper is arranged as follows.
In Sec.~\ref{sec:protocol}, we present the procedures of our general SNS protocol with asymmetric source parameters.
In Sec.~\ref{sec:proof}, we give a security proof of our protocol through three virtual protocols and their reductions.
We show the results of numerical simulation of the general SNS protocol compared with the original SNS protocol in Sec.~\ref{sec:simulation}.
The article ends with some concluding remarks in Sec.~\ref{sec:conclusion}.
We give the formulas for key rate calculation in the appendix.

\section{General SNS protocol with asymmetric source parameters}\label{sec:protocol}
\begin{figure}
    \includegraphics[width=200pt]{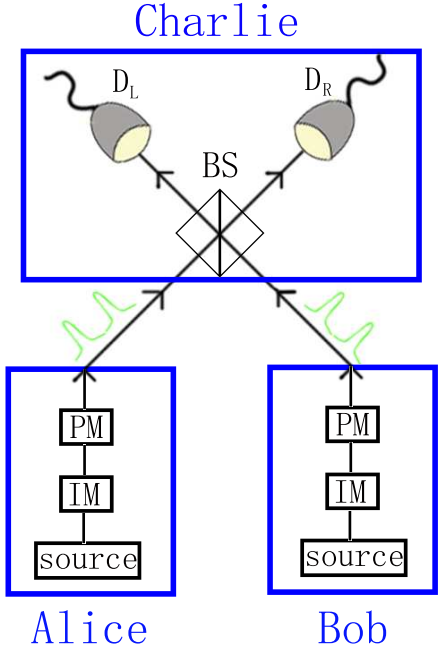}
    \caption{A schematic of the setup for the general SNS protocol. IM: intensity modulator; PM: phase modulator; BS: beam-splitter; $D_L$ \& $D_R$: single-photon detector in the measurement station of Charlie.}\label{fig:sketch}
    \end{figure}

A schematic of our general SNS protocol is shown in Figure~\ref{fig:sketch}.
The two legitimate users, Alice and Bob, independently send coherent pulses and vacuum pulses to an untrusted third party (UTP), Charlie.
Charlie takes compensation to the pulses, measures them, and announces the measurement results.
Then Alice and Bob distill the final key from a set of the pulses according to the announced data.
The details of the protocol are shown as follows.

{\bf Step 1.} In each time window $i$, {\em they} (Alice and Bob) independently decide whether this is a decoy window or a signal window.
In her (his) decoy window, she (he) randomly chooses one of a few states $\rho_{Ak}$ ($\rho_{Bk}$), for $k=0,1,2,\dots$, to send out a decoy pulse to Charlie where $\rho_{A0}=\rho_{B0}=\oprod{0}{0}$ are the vacuum states and $\rho_{Ak}$ ($\rho_{Bk}$), $k>0$, are coherent states $\ket{\sqrt{\mu_{Ak}} e^{\mathbf{i}\delta_{Ai}+\mathbf{i}\gamma_{Ai}}}$ ($\ket{\sqrt{\mu_{Bk}} e^{\mathbf{i}\delta_{Bi}+\mathbf{i}\gamma_{Bi}}}$).
(In this paper, we denote the imaginary unit as $\mathbf{i}$.)
In her (his) signal window, she (he) decides to send out to Charlie a signal pulse in the state $\ket{\sqrt{\mu_{A}^\prime} e^{\mathbf{i}\delta_{Ai}+\mathbf{i}\gamma_{Ai}}}$ ($\ket{\sqrt{\mu_{B}^\prime} e^{\mathbf{i}\delta_{Bi}+\mathbf{i}\gamma_{Bi}}}$) and puts down a bit value 1 (0) by probability $\epsilon_A$ ($\epsilon_B$), or decides not to send it out and puts down a bit value 0 (1) by probability $1-\epsilon_A$ ($1-\epsilon_B$).
Here, $\delta_{Ai}$, $\delta_{Bi}$, $\gamma_{Ai}$, and $\gamma_{Bi}$ are random phase.
The global phases $\gamma_{Ai}$ and $\gamma_{Bi}$ can be any reference phase and known by anyone.
The private phase $\delta_{Ai}$ ($\delta_{Bi}$) is a random phase taken by Alice (Bob) secretly.
Besides, we request the following mathematical constraint for source parameters:
\begin{equation}\label{equ:protocol1}
   \frac{\mu_{Ak}}{ \mu_{Bk}} = \frac{\epsilon_A(1-\epsilon_B) \mu_A^\prime  e^{-\mu_A^\prime}}{\epsilon_B(1-\epsilon_A)  \mu_B^\prime e^{-\mu_B^\prime}}
\end{equation}
for each $k>0$.

As the major result of this work, this constraint guarantees the security of the general SNS protocol with asymmetric source parameters for Alice and Bob.
The real protocol here is actually same with that in ref.~\cite{wang2018twin} except for this mathematical constraint and the virtual protocols for security proof will use different types of entangled states.

For ease of presentation, we will omit the subscript $i$ if it doesn't cause any confusion.
But keep in mind that all $\delta_A$, $\delta_B$, $\gamma_A$, and $\gamma_B$ are chosen differently in different time windows.

{\bf Step 2.} Charlie is supposed to measure all twin fields with a beam splitter after taking phase compensation and announce the measurement outcome.

Note: Charlie is expected to remove the phase $\gamma_{A}$ and $\gamma_{B}$ by phase compensation.
His action only affects the key rate, but has not influence on the security of the protocol.

{\bf Step 3.} {\em They} announce each one's decoy windows and signal windows.
And {\em they} announce the intensities of the decoy pulses and the private phases $\delta_{A}$ and $\delta_{B}$ in each decoy window.

{\em Definition: }We define a $Z$-window when both of {\em them} determine signal windows and an $X$-window when both of {\em them} determine decoy windows.
We define an {\em effective event} when one and only one Charlie's detector clicks, and the corresponding window is called an {\em effective window}.

Given that $\delta_A$ ($\delta_B$) is randomized, whenever Alice (Bob) sends a coherent state with intensity $\mu_A^\prime$ ($\mu_B^\prime$), it can be equivalently regarded as a density matrix $\sum_{k=0}^\infty\frac{e^{-\mu_A^\prime}\mu_A^{\prime k}}{k!}\oprod{k}{k}$ ($\sum_{k=0}^\infty\frac{e^{-\mu_B^\prime}\mu_B^{\prime k}}{k!}\oprod{k}{k}$), which is a classical mixture of different photon number states only.
Hence among all $Z$-windows, we can define a set of $Z_1$-windows, in which one and only one of {\em them} decides to send and she (he) actually sends a single-photon state.
{\em They} don't know which time window is a $Z_1$-window, but {\em they} can calculate the number of $Z_1$-windows in an experiment.

Among all $X$-windows, we define a set of $\tilde{X}$-windows, in which {\em they} choose the intensities $\mu_{Ak}$ and $\mu_{Bk}$ with the same $k$ and the phase shifts $\delta_A$ and $\delta_B$ satisfy the restriction
\begin{equation}\label{equ:restriction}
    1 - |\cos (\delta_{A} - \delta_{B} + \Delta\varphi)| \le |\lambda|.
\end{equation}
Here the values $\Delta\varphi$ and $\lambda$ are some values determined by Alice and Bob according to the result of channel test and calibration in the experiment to obtain a satisfactory key rate and the value of $\Delta\varphi$ can be different from time to time.
In this paper, we will set $\Delta\varphi=0$ for presentation simplicity, i.e.
\begin{equation}\label{equ:criterion}
    1 - |\cos (\delta_{A} - \delta_{B})| \le |\lambda|,
\end{equation}
but this doesn't affect the validity of the security proof if we use Eq.(\ref{equ:restriction}) for the post selection.
{\em They} keep the data of the effective events and discard all the others.

{\bf Step 4.} {\em They} randomly choose some events from the effective $Z$-windows to do the error test.
A bit-flip error occurs when Alice's bit value is different from Bob's in a $Z$-window.
{\em They} discard the test bits, and the remaining events from effective $Z$-windows will be distilled for the final key.

{\bf Step 5.} Based on the measurement outcome in effective $X$-windows, {\em they} calculate $n_1$, the number of effective events in $Z_1$-windows.
Based on the measurement outcome and the announced values of $\delta_A$ and $\delta_B$ in effective $\tilde{X}$-windows, {\em they} calculate $e_1^{ph}$, the phase-flip error rate of effective events in $Z_1$-windows.

{\em Note:} In effective $\tilde{X}$-windows, an error occurs when \\
(1)the left detector clicks and $\cos(\delta_A-\delta_B)<0$ \\
(2)the right detector clicks and $\cos(\delta_A-\delta_B)>0$.\\
Given this definition, {\em they} can observe the error rates in $\tilde{X}$-windows for each intensities of input light.
With this, {\em they} can estimate $e_1^{ph}$ through the decoy state analysis which requests them to observe the counting rates of various intensities of input light.
As proved in ref.~\cite{wang2018twin}, decoy state method can applied to our protocol as if the phases $\delta_A$ and $\delta_B$ were not announced.
In the asymptotic case that there are decoy states with infinite different intensities, {\em they} can obtain the exact value of $e_1^{ph}$. In the case that there are decoy states with finite different intensities, {\em they} can obtain the upper bound of $e_1^{ph}$.

{\em Note:} The Appendix shows the four-intensity method of this protocol. In this case, the formulas for $n_1$ and $e_1^{ph}$ are given in Eqs.(\ref{equ:s1Z})-(\ref{equ:e1ph}).

{\bf Step 6.} {\em They} perform the post-processing and obtain the final key with length
\begin{equation}\label{equ:length}
    N_f = n_1 [1 - H(e_1^{ph})] - f n_t H(E_Z)
\end{equation}
where $f$ is the correction efficiency, $n_t$ is the number of effective $Z$-windows, and $E_Z$ is the bit-flip error rate in effective $Z$-windows. Details for calculating the length of the final key (or the key rate) with four-intensity decoy-state method are presented in the Appendix.

{\em Note:} If we set $\epsilon_A=\epsilon_B$ and $\mu_A^\prime=\mu_B^\prime$, this protocol is actually the same as the original SNS protocol in ref.~\cite{wang2018twin}.

\section{Security proof with virtual protocols and reduction}\label{sec:proof}

\subsection{Introduction of the ancillary photons and the extended states}
Similarly to the security proof in Ref.~\cite{wang2018twin}, we use the idea of entanglement distillation with ancillary photons to proof the security of our protocol.

Image that in a $Z$-window, if Alice (Bob) decides to send a coherent state $\rho_A$ ($\rho_B$) to Charlie, she (he) puts down a local ancillary qubit in the state $\ket{1}$ ($\ket{0}$), and if Alice (Bob) decides not to send, she (he) puts down a local ancillary qubit in the state $\ket{0}$ ($\ket{1}$).
To Alice (Bob), the state $\ket{1}$ corresponds to the bit value 1 (0), and  the state $\ket{0}$ corresponds to the bit value 0 (1).
We define subspace $\mathcal{T}$ for the subspace of the sent-out states and $\mathcal{A}n$ for the subspace of the local ancillary states.
Therefore, the extended state in the complex space $\mathcal{T}\otimes\mathcal{A}n$ in the $Z$-window can be written as
\begin{equation}\label{equ:Omega}
\begin{split}
    \Omega &= \epsilon_A \epsilon_B (\rho_A \tilde{\otimes} \rho_B) \otimes \oprod{11}{11} \\
    &+ \epsilon_A (1-\epsilon_B) (\rho_A \tilde{\otimes} \oprod{0}{0}) \otimes \oprod{10}{10} \\
    &+ (1-\epsilon_A) \epsilon_B (\oprod{0}{0} \tilde{\otimes} \rho_B) \otimes \oprod{01}{01} \\
    &+ (1-\epsilon_A) (1-\epsilon_B) (\oprod{0}{0} \tilde{\otimes} \oprod{0}{0}) \otimes \oprod{00}{00}.
\end{split}
\end{equation}
Here both symbols $\otimes$ and $\tilde{\otimes}$ are for a tensor product, and $\tilde{\otimes}$ is the tensor product inside $\mathcal{T}$, and $\otimes$ is the tensor product between $\mathcal{T}$ and $\mathcal{A}n$.
The states on the left side of $\otimes$ are in $\mathcal{T}$ and we name them {\em real-photon states}.
The states on the right side of $\otimes$ are in $\mathcal{A}n$ and we name them {\em ancillary-photon states}.
Since the private phases $\delta_A$ and $\delta_B$ in $Z$-windows are kept secret all the time, the coherent states $\rho_A$ and $\rho_B$ are actually phase-randomized coherent states, whose density matrices are
\begin{equation}\label{equ:rhoAB}
    \rho_k = \sum_{n=0}^\infty \frac{e^{-\mu^\prime_k}\mu_k^{\prime n}}{n!}\oprod{n}{n}= \mu_k^{\prime}e^{-\mu^\prime_k}\oprod{1}{1} + (1-\mu_k^{\prime}e^{-\mu^\prime_k})\bar{\rho}_k,
\end{equation}
with $k=A,B$ and
\begin{equation}\label{equ:rhobar}
    \bar{\rho}_k = \frac{1}{1-\mu_k^{\prime}e^{-\mu^\prime_k}} \sum_{n\neq1} \frac{e^{-\mu^\prime_k}\mu_k^{\prime n}}{n!}\oprod{n}{n}.
\end{equation}
So the extended state in the $Z$-window can be written in another form:
\begin{equation}\label{equ:Omeganew}
    \Omega=\sum_{r=1}^4 q_r \Omega_r
\end{equation}
and
\begin{equation}\label{equ:Omegar}
\begin{split}
    \Omega_1 =& C_1 [\epsilon_A(1-\epsilon_B) \mu_A^\prime e^{-\mu_A^\prime} \oprod{10}{10} \otimes \oprod{10}{10} \\
    & + \epsilon_B(1-\epsilon_A) \mu_B^\prime e^{-\mu_B^\prime} \oprod{01}{01} \otimes \oprod{01}{01}] \\
    \Omega_2 =& C_2 [\epsilon_A(1-\epsilon_B) (1-\mu_A^\prime e^{-\mu_A^\prime}) (\bar\rho_A \tilde\otimes \oprod{0}{0}) \otimes \oprod{10}{10} \\
    & + \epsilon_B(1-\epsilon_A) (1-\mu_B^\prime e^{-\mu_B^\prime}) (\oprod{0}{0} \tilde\otimes \bar\rho_B) \otimes \oprod{01}{01}] \\
    \Omega_3 =& \oprod{00}{00} \otimes \oprod{00}{00} \\
    \Omega_4 =& (\rho_A \tilde\otimes \rho_B) \otimes \oprod{11}{11}
\end{split}
\end{equation}
where $C_1$ and $C_2$ are some normalization factors.
With the condition in Eq. (\ref{equ:protocol1}), $\Omega_1$, the target state we used to prove the security, can be written as
\begin{equation}\label{equ:Omega1}
\begin{split}
    \Omega_1 =& C^2 (\mu_{A1}  \oprod{10}{10} \otimes \oprod{10}{10} + \mu_{B1}  \oprod{01}{01} \otimes \oprod{01}{01})
\end{split}
\end{equation}
with
\begin{equation}\label{equ:C}
    C = 1 / \sqrt{\mu_{A1} + \mu_{B1}}.
\end{equation}

In $X$-windows, the two-mode states sent by {\em them} are
\begin{equation}\label{equ:rhoX}
    \rho_{Xk} = \oprod{\beta_k}{\beta_k}
\end{equation}
where
\begin{equation}\label{equ:betak}
    \ket{\beta_k} = \ket{\sqrt{\mu_{Ak}} e^{\mathbf{i}\delta_{A}+\mathbf{i}\gamma_{A}}} \ket{\sqrt{\mu_{Bk}} e^{\mathbf{i}\delta_{B}+\mathbf{i}\gamma_{B}}}.
\end{equation}

In our protocol, the states in the $Z$-windows are actually classical mixtures of $\Omega_1$, $\Omega_2$, $\Omega_3$, and $\Omega_4$.
In the security proof, we first show the security of the protocol with only the state $\Omega_1$ and then show the security of the protocol with $\Omega$ by the tagged model~\cite{inamori2007unconditional,gottesman2004security}.


\subsection{Virtual Protocol 1}
{\em Definition:}
We have define an {\em effective event} in the protocol.
An {\em effective ancillary photon} is an ancillary photon corresponding to an effective event.

\subsubsection{Preparation stage}
For each time window $i$, {\em they} preshare the classical information about whether this time window is an $X$-window or $Z$-window.
{\em They} preshare an extended state
\begin{equation}\label{equ:Omega0}
    \Omega_{0} = \oprod{\Psi}{\Psi}
\end{equation}
where
\begin{equation}\label{equ:Psi}
    \ket{\Psi} = C (\sqrt{\mu_{A1}} e^{\mathbf{i}\gamma_{A}} \ket{10} \otimes \ket{10} + \sqrt{\mu_{B1}} e^{\mathbf{i}\gamma_{B}} \ket{01} \otimes \ket{01})
\end{equation}
and $\gamma_{A}$ and $\gamma_{B}$ are announced publicly. (Remind that $\Omega_0$ and $\ket\Psi$ vary in different time windows.)

In the time window $i$ which is a $Z$-window, through discussion by a secret channel, Alice chooses a random phase $\delta_{A}$ and Bob chooses a random phase $\delta_{B}$, which satisfy the restriction Eq.(\ref{equ:criterion}).
Then {\em they} take phase shifts $\delta_A$ and $\delta_B$ to their own real photons, respectively.
In the time window $i$ which is a $X$-window, {\em they} take random and independent phase shifts $\delta_A$ and $\delta_B$ to their own real photons, respectively.
After the phase shifts, the extended state changes into
\begin{equation}\label{equ:OmegaZX}
    \Omega_{Z} = \oprod{\Psi^\prime}{\Psi^\prime},\ \Omega_{X} = \oprod{\Psi^\prime}{\Psi^\prime}
\end{equation}
and
\begin{equation}\label{equ:Psi'}
\begin{split}
    \ket{\Psi^\prime} = C (&\sqrt{\mu_{A1}} e^{\mathbf{i}\gamma_{A} + \mathbf{i}\delta_{A}} \ket{10} \otimes \ket{10} \\
    &+ \sqrt{\mu_{B1}} e^{\mathbf{i}\gamma_{B} + \mathbf{i}\delta_{B}} \ket{01} \otimes \ket{01})
\end{split}
\end{equation}

Among all $X$-windows, we define a set of $\tilde{X}$-windows, in which the phase shifts $\delta_A$ and $\delta_B$ satisfy the restriction Eq.(\ref{equ:criterion}).
The states in $Z$-windows are not identical to those in $X$-windows, but they are identical to those in $\tilde{X}$-windows.

Besides, we define real-photon states $\ket{\chi^0}$ and $\ket{\chi^1}$ for any time window:
\begin{equation}\label{equ:chi0}
\begin{split}
    \ket{\chi^0} =& C (\sqrt{\mu_{A1}} e^{\mathbf{i}\gamma_{A} + \mathbf{i}\delta_{A}} \ket{10} + \sqrt{\mu_{B1}} e^{\mathbf{i}\gamma_{B} + \mathbf{i}\delta_{B}} \ket{01}) \\
    \ket{\chi^1} =& C (\sqrt{\mu_{A1}} e^{\mathbf{i}\gamma_{A} + \mathbf{i}\delta_{A}} \ket{10} - \sqrt{\mu_{B1}} e^{\mathbf{i}\gamma_{B} + \mathbf{i}\delta_{B}} \ket{01}) \\
    &\text{if}\quad \cos(\delta_A-\delta_B) \ge 0
\end{split}
\end{equation}
or
\begin{equation}\label{equ:chi1}
\begin{split}
    \ket{\chi^0} =& C (\sqrt{\mu_{A1}} e^{\mathbf{i}\gamma_{A} + \mathbf{i}\delta_{A}} \ket{10} - \sqrt{\mu_{B1}} e^{\mathbf{i}\gamma_{B} + \mathbf{i}\delta_{B}} \ket{01}) \\
    \ket{\chi^1} =& C (\sqrt{\mu_{A1}} e^{\mathbf{i}\gamma_{A} + \mathbf{i}\delta_{A}} \ket{10} + \sqrt{\mu_{B1}} e^{\mathbf{i}\gamma_{B} + \mathbf{i}\delta_{B}} \ket{01}) \\
    &\text{if}\quad \cos(\delta_A-\delta_B) < 0
\end{split}
\end{equation}

\subsubsection{Protocol}
{\bf V1-1: }At any $Z$-window ($X$-window), {\em they} send the real photons of $\Omega_Z$ ($\Omega_X$) defined in Eq. (\ref{equ:OmegaZX}) to Charlie and keep the ancillary photons locally.

{\bf V1-2: }Charlie measures the real photons from Alice and Bob with a beam splitter after taking phase compensation according to the strong reference lights with phases $\gamma_A$ and $\gamma_B$.
He announces his measurement outcome and then {\em they} announce the values of $\delta_A$ and $\delta_B$ of all $X$-windows.
With the preshared information of $X$- and $Z$-windows, the measurement outcome, and the announced values of $\delta_A$ and $\delta_B$, {\em they} can obtain effective $Z$-windows and effective $\tilde{X}$-windows.
The data of other time windows will be discarded.

{\em Definition: }After Step V1-2, the remaining effective events can be divided into eight subsets according to the window information ($Z$-window or $X$-window), the clicking detector (the left $L$ or the right $R$) and the sign of $\cos(\delta_A-\delta_B)$ (positive $+$ or negative $-$).
These subsets is labeled as $\Gamma_{(a,d)}$ where $\Gamma=\tilde{X},Z$, $a=+,-$, and $d=L,R$.
For example, the subset $Z_{(-,R)}$ is the set of effective $Z$-windows when the right detector clicks and $\cos(\delta_A-\delta_B)<0$.
Correspondingly, the effective ancillary photons can be divide into eight subsets labeled as $\mathcal{A}_{\Gamma_{(a,d)}}$.

{\bf V1-3: }{\em They} check the phase-flip error rate $E_{(a,d)}$ of the set $\mathcal{A}_{\tilde{X}_{(a,d)}}$, where $a=+,-$ and $d=L,R$.
Since the effective $Z$-windows are identical to the effective $\tilde{X}$-windows, the phase-flip error rate of $\mathcal{A}_{\tilde{X}_{(a,d)}}$ should be the same as that of $\mathcal{A}_{Z_{(a,d)}}$ (asymptotically).

{\em Note: }The state $\ket{\Psi^\prime}$ in Eq.(\ref{equ:Psi'}) can be written in another form:
\begin{equation}\label{equ:Psi'1}
    \ket{\Psi^\prime} = \frac{1}{\sqrt2} (\ket{\chi^0}\otimes\ket{\Phi^0} + \ket{\chi^1}\otimes\ket{\Phi^1}),\ \text{if }a=+
\end{equation}
or
\begin{equation}\label{equ:Psi'2}
    \ket{\Psi^\prime} = \frac{1}{\sqrt2} (\ket{\chi^1}\otimes\ket{\Phi^0} + \ket{\chi^0}\otimes\ket{\Phi^1}),\ \text{if }a=-
\end{equation}
where $\ket{\Phi^k},k=0,1$ are two-mode states of ancillary photons:
\begin{equation}\label{equ:Phi}
    \ket{\Phi^0} = \frac{1}{\sqrt2} (\ket{10}+\ket{01}),\ket{\Phi^1} = \frac{1}{\sqrt2} (\ket{10}-\ket{01})
\end{equation}
To obtain the phase-flip error rate $E_{(a,d)}$ of the set $\mathcal{A}_{\tilde{X}_{(a,d)}}$, each ancillary photon of this set is measured in the basis $\{\ket{\Phi^0},\ket{\Phi^1}\}$ and there are $n_{(a,d)}^{(0)}$ outcomes of $\oprod{\Phi^0}{\Phi^0}$ and $n_{(a,d)}^{(1)}$ outcomes of $\oprod{\Phi^1}{\Phi^1}$.
The phase-flip error rate $E_{(a,d)}$ of $\mathcal{A}_{\tilde{X}_{(a,d)}}$ is define as:
\begin{equation}\label{equ:Ead}
    E_{(a,d)} = \frac{\min \big(n_{(a,d)}^{(0)},n_{(a,d)}^{(1)}\big)}{n_{(a,d)}}
\end{equation}
where $n_{(a,d)}=n_{(a,d)}^{(0)}+n_{(a,d)}^{(1)}$ is the number of the ancillary photons in $\mathcal{A}_{\tilde{X}_{(a,d)}}$.

{\bf V1-4: }With the estimated value of $E_{(a,d)}$, {\em they} can purify the ancillary photons in $\mathcal{A}_{Z_{(a,d)}}$ with $(a,d)=(+,R),(+,L),(-,R),(-,L)$ separately.
After purification, {\em they} obtain ancillary photons in (almost 100\%) pure single-photon entangled states $\ket{\Phi^0}$ (or $\ket{\Phi^1}$).
Then {\em they} perform local measurement on their own ancillary photons and obtain the final key.
Alice (Bob) puts down a bit value 0 (1) or 1 (0) when her (his) measurement outcome is $\oprod{0}{0}$ or $\oprod{1}{1}$ .

{\em Note 1-Security: }The security of the final key is based on the faithfulness of the purification.
If {\em they} estimate the error rate $E_{(a,d)}$ of a set of ancillary photons exactly, {\em they} can purify these ancillary photons to get pure entangled photons.
Although Charlie measured the real photons and {\em they} selected the set of ancillary photons based on his announced measurement outcomes, {\em they} check the phase-flip error rate of these photons by themselves.
Since the extended states of effective $\tilde{X}$-windows are identical to those of effective $Z$-windows, the phase-flip error rate of $\mathcal{A}_{\tilde{X}_{(a,d)}}$ is exactly the same as that of $\mathcal{A}_{Z_{(a,d)}}$ statistically.
{\em They} can obtain the phase-flip error rate in $\mathcal{A}_{Z_{(a,d)}}$ by testing the ancillary photon in $\mathcal{A}_{\tilde{X}_{(a,d)}}$ and then perform the purification to $\mathcal{A}_{Z_{(a,d)}}$.
{\bf As a result, the security of the key doesn't rely on Charlie's honesty}.

{\em Note 2-Estimation of the phase-flip error rate: }According to the definition of $E_{(a,d)}$ in Eq.(\ref{equ:Ead}), {\em they} have to measure the ancillary photons in the basis $\{\ket{\Phi^0},\ket{\Phi^1}\}$. It's easy to prove that {\em they} can perform local measurement in the basis $\{\ket{\pm x}=(\ket0\pm\ket1)/\sqrt2\}$ and check the parity of the outcome instead of measuring in the basis $\{\ket{\Phi^0},\ket{\Phi^1}\}$. Explicitly, their outcome with even parity ($\ket{x+}\ket{x+}$ or $\ket{x-}\ket{x-}$) corresponds to the outcome $\ket{\Phi^0}$ and that with odd parity ($\ket{x+}\ket{x-}$ or $\ket{x-}\ket{x+}$) corresponds to the outcome $\ket{\Phi^1}$.

{\em Note 3-Reduction of the preshare states in $X$-windows: }Since measurement in the basis $\{\ket{\pm x}\}$ is a local operation on the ancillary photons, it makes no difference whether {\em they} measure their ancillary photons after sending the real photons or before that.
So they can measure the ancillary photons before the protocol starts, and label this time window an $X_0$-window if the outcome is even parity, or label it an $X_1$-window if the outcome is odd parity.
Then {\em they} prepare and send the real photon in the state
\begin{equation}\label{equ:chi+}
    \ket{\chi^+} = C (\sqrt{\mu_{A1}} e^{\mathbf{i}\gamma_{A} + \mathbf{i}\delta_{A}} \ket{10} + \sqrt{\mu_{B1}} e^{\mathbf{i}\gamma_{B} + \mathbf{i}\delta_{B}} \ket{01})
\end{equation}
in an $X_0$-window, or prepare and send that in the state
\begin{equation}\label{equ:chi-}
    \ket{\chi^-} = C (\sqrt{\mu_{A1}} e^{\mathbf{i}\gamma_{A} + \mathbf{i}\delta_{A}} \ket{10} - \sqrt{\mu_{B1}} e^{\mathbf{i}\gamma_{B} + \mathbf{i}\delta_{B}} \ket{01})
\end{equation}
in an $X_1$-window.

Alternatively, {\em they} can start with the information of $X_0$-windows and $X_1$-windows and the states in Eqs. (\ref{equ:chi+})(\ref{equ:chi-}).
{\em They} prepare and send real photons in the state $\ket{\chi^+}$ in $X_0$-windows or in the state $\ket{\chi^-}$ in $X_1$-windows.
In this way, the ancillary photons in $X$-windows in the above virtual protocol are not necessary, and the formula of phase-flip error rate should be changed correspondingly.
We introduce a symbol $\tilde{X}_{(a,b,d)}$ for the set of effective time windows, which satisfy the restriction in Eq.(\ref{equ:criterion}), with joint events $a$, $b$, $d$ where

Event $a$ ($a=+,-$): the sign of $\cos(\delta_A-\delta_B)$.

Event $b$ ($b=0,1$): this times window is an $X_b$-window.

Event $d$ ($d=L,R$): the $d$-detector clicks and the other doesn't click.

And we also introduce $n_{\tilde{X}_{(a,b,d)}}$ for the number of time windows in the set $\tilde{X}_{(a,b,d)}$.
Therefore, we have
\begin{equation}\label{equ:NX}
    n^{(b)}_{(a,d)} = N_{\tilde{X}_{(a,b,d)}}
\end{equation}
and
\begin{equation}\label{equ:Ead'}
    E_{(a,d)} = \frac{\min (N_{\tilde{X}_{(a,0,d)}},N_{\tilde{X}_{(a,1,d)}})}{n_{(a,d)}}
\end{equation}
This reduction of $X$-windows leads to the  Virtual Protocol 2.

\subsection{Virtual Protocol 2}
\subsubsection{Preparation stage}
For each time window $i$, {\em they} preshare the classical information about whether this time window is an $X_0$-window, an $X_1$-window or a $Z$-window.

{\em They}  preshare an extended state $\Omega_Z$ in Eq.(\ref{equ:OmegaZX}) for $Z$-windows, an real-photon state $\ket{\chi^+}$ for $X_0$-windows and an real state $\ket{\chi^-}$ for $X_1$-windows.

\subsubsection{Protocol}
{\bf V2-1: }At any $Z$-window, {\em they} send out the real photons of $\Omega_Z$ to Charlie and keep the ancillary photons locally. At any $X_0$-window ($X_1$-window), {\em they} send $\ket{\chi^+}$ ($\ket{\chi^-}$) to Charlie.

{\bf V2-2: }Charlie measures the real photons from Alice and Bob with a beam splitter after taking phase compensation according to the strong reference lights with phases $\gamma_A$ and $\gamma_B$.
He announces his measurement outcome and then {\em they} announce the values of $\delta_A$ and $\delta_B$ of all $X$-windows.

{\bf V2-3: }{\em They} check the phase-flip error rate $E_{(a,d)}$ by the set $\tilde{X}_{(a,0,d)}$ and $\tilde{X}_{(a,1,d)}$, where $a=+,-$ and $d=L,R$.

{\bf V2-4: }{\em They} purify the ancillary photons in $\mathcal{A}_{Z_{(a,d)}}$ with $(a,d)=(+,R),(+,L),(-,R),(-,L)$ separately with the estimated value of $E_{(a,d)}$.
Then {\em they} perform local measurement on their own ancillary photons and obtain the final key.

{\em Note: Reduction of preshared states in $X$-windows.}
\\{\em Reduction 1: }The real-photon states with $a=+$ ($a=-$) in $X_0$-windows are actually identical to those with $a=-$ ($a=+$) in $X_1$-windows, e.g.
\begin{equation}
    \rho_{(+,0)} = \rho_{(-,1)},\rho_{(-,0)} = \rho_{(+,1)}.
\end{equation}
So we can conclude that $N_{\tilde{X}_{(a,0,d)}}=N_{\tilde{X}_{(\bar{a},1,d)}}$, where $\bar{a}$ stands for the opposite sign of $a$.
This means that all the values $N_{\tilde{X}_{(a,1,d)}}$ in the phase-flip error rate in Eq.(\ref{equ:Ead'}) can be replaced by $N_{\tilde{X}_{(\bar{a},0,d)}}$ so that $X_1$-windows are not necessary.
{\em They} can just use the data from $X_0$-windows to estimate the phase-flip error rate, and no one else will find any difference.
Therefore, {\em they} use only $X_0$-windows and send only the state $\ket{\chi^+}$ in $X$-windows.
The number of effective events from the state $\ket{\chi^+}$ and the joint events $a$, $d$ (with $a=+,-;d=L,R$) is denoted as $n_{\tilde{X}_{(a,d)}}$.
The formula of the phase-flip error rate should be changed into:
\begin{equation}\label{equ:Ead''}
    E_{(a,d)} = \frac{\min (n_{\tilde{X}_{(a,d)}},n_{\tilde{X}_{(\bar{a},d)}})}{n_{(a,d)}}
\end{equation}
where the formula for $n_{(a,d)}$ should be changed into $n_{(a,d)} = n_{\tilde{X}_{(a,d)}}+n_{\tilde{X}_{(\bar{a},d)}}$.
\\{\em Reduction 2: }All the effective ancillary photons in $Z$-windows can be purified in one batch. The phase-flip error rate is
\begin{equation}
\begin{split}
    E^{ph} = \frac{\sum_{a,d} \min (n_{\tilde{X}_{(a,d)}},n_{\tilde{X}_{(\bar{a},d)}})}{\sum_{a,d} n_{(a,d)}} \\
    =\frac{2\sum_{d} \min (n_{\tilde{X}_{(+,d)}},n_{\tilde{X}_{(-,d)}})}{2 n_1}
\end{split}
\end{equation}
where $n_1=n_{\tilde{X}_{(+,L)}}+n_{\tilde{X}_{(+,R)}}+n_{\tilde{X}_{(-,L)}}+n_{\tilde{X}_{(-,R)}}$ is the total number of effective events in $\tilde{X}$-windows.
Using the relations that $n_{\tilde{X}_{(-,L)}} \ge \min(n_{\tilde{X}_{(+,L)}},n_{\tilde{X}_{(-,L)}})$ and $n_{\tilde{X}_{(+,R)}} \ge \min(n_{\tilde{X}_{(+,R)}},n_{\tilde{X}_{(-,R)}})$, the phase-flip error rate can be bounded by
\begin{equation}\label{equ:Eph}
    E^{ph} \le \frac{n_{\tilde{X}_{(-,L)}} + n_{\tilde{X}_{(+,R)}}}{n_1}.
\end{equation}
In this formula for the phase-flip error rate, we only need the total number of effective events in $\tilde{X}$-windows and the number of these two kinds of effective events:
\\1. the left detector clicks and $\cos(\delta_A-\delta_B)<0$;
\\2. the right detector clicks and $\cos(\delta_A-\delta_B)\ge0$.
\\Therefore, we can define these two kinds of effective events as {\em error events} and the corresponding time windows are defined as {\em error windows}. If {\em they} set the value of $\lambda$ small enough and Charlie perform the compensation honestly, {\em they} may get few error events so that the phase-flip error rate will be quite low and the key rate will be high.
\\
\\{\em Reduction 3: }The density matrix in Eq.(\ref{equ:rhoX}) with randomized $\delta_A$ and $\delta_B$ can be written as the classical mixture of a set of states $\{\ket{\psi_l^{(k)}}\}$:
\begin{equation}\label{equ:rhoX'}
    \rho_{Xk} = \sum_l p_l^{(k)} \oprod{\psi_l^{(k)}}{\psi_l^{(k)}}
\end{equation}
with
\begin{equation}\label{equ:psik}
\begin{split}
    \ket{\psi_l^{(k)}} = D_l^{(k)} \sum_{n=0}^l \frac{(\sqrt{\mu_{Ak}} e^{\mathbf{i}\delta_{A} + \mathbf{i}\gamma_{A}})^n}{\sqrt{n!}} \frac{(\sqrt{\mu_{Bk}} e^{\mathbf{i}\delta_{B} + \mathbf{i}\gamma_{B}})^{l-n}}{\sqrt{(l-n)!}} \\
    \ket{n,l-n}
\end{split}
\end{equation}
where $D_l^{(k)}$ are some normalization factors and $\ket{\psi_1^{(k)}}$ is exactly $\ket{\chi^+}$ when the condition in Eq.(\ref{equ:protocol1}) is satisfied.
So {\em they} don't need to preshare the state $\ket{\chi^+}$.
{\em They} can send the phase-randomized coherent state
$\ket{\sqrt{\mu_{Ak}} e^{\mathbf{i}\delta_{A} + \mathbf{i}\gamma_{A}}}$ and $\ket{\sqrt{\mu_{Bk}} e^{\mathbf{i}\delta_{B} + \mathbf{i}\gamma_{B}}}$ to Charlie in $X$-windows, and then use the decoy-state method to estimate the bound of the phase-flip error rate of $\ket{\chi^+}$, $e_1^{ph}$.
\\{\em Note:} If some decoy states with intensities $\mu_{Am}$ and $\mu_{Bm}$ are not used to estimate $e_1^{ph}$, these intensities don't have to satisfy the constrain in Eq.(\ref{equ:protocol1}).

\subsection{Virtual Protocol 3}
\subsubsection{Preparation stage}
For each time window $i$, {\em they} preshare the classical information about whether this time window is an $X$ window or a $Z$-window.

{\em They}  preshare an extended state $\Omega_Z$ in Eq.(\ref{equ:OmegaZX}) for $Z$-windows.

\subsubsection{Protocol}
{\bf V3-1: }At any $Z$-window, {\em they} send out the real photons of $\Omega_Z$ to Charlie and keep the ancillary photons locally.
At any $X$-window, Alice (Bob) sends a coherent state $\ket{\sqrt{\mu_{Ak}} e^{\mathbf{i}\delta_{A} + \mathbf{i}\gamma_{A}}}$ ($\ket{\sqrt{\mu_{Bk}} e^{\mathbf{i}\delta_{B} + \mathbf{i}\gamma_{B}}}$) with random $\delta_A$ and $\gamma_A$ ($\delta_B$ and $\gamma_B$) to Charlie.

{\bf V3-2: }Charlie measures the real photons from Alice and Bob with a beam splitter after taking phase compensation according to the strong reference lights with phases $\gamma_A$ and $\gamma_B$.
He announces his measurement outcome and then {\em they} announce the values of $\delta_A$ and $\delta_B$ of all $X$-windows.

{\bf V3-3: }{\em They} apply decoy-state method with the data of the effective $\tilde{X}$-windows to estimate the phase-flip error rate $e_1^{ph}$.

{\bf V3-4: }{\em They} purify the ancillary photons in the effective $Z$-windows with the estimated value of $e_1^{ph}$.
Then {\em they} perform local measurement on their own ancillary photons and obtain the final key.

{\em Note 1: Reduction of preshared states in $Z$-windows.}
\\{\em Reduction 1: }The state $\Omega_Z$ with the restriction Eq.(\ref{equ:criterion}) is identical to $\Omega_Z$ without the restriction Eq.(\ref{equ:criterion}).
If we regard all $Z$-windows as a whole, the condition in Eq.(\ref{equ:criterion}) can be loosen, which means that the phase shifts $\delta_A$ and $\delta_B$ to the real photons can be randomized in the range $[0,2\pi)$ independently.
In this way, {\em they} don't need any mutual information about the phase shifts $\delta_A$ and $\delta_B$.
\\
\\{\em Reduction 2: }The process that {\em they} purify the effective ancillary photons in $Z$-windows and then perform local measurement on them is equivalent to the process that {\em they} measure these ancillary photons in the photon-number basis and then do classical distillation to the classical data, which is called quasipurification~\cite{shor2000simple}.
In the latter process, the state in $Z$-window is
\begin{equation}\label{equ:OmegaZ'}
    \Omega_Z^\prime = C^2 (\mu_{A1} \oprod{10}{10} \otimes \oprod{10}{10} + \mu_{B1} \oprod{01}{01} \otimes \oprod{01}{01}),
\end{equation}
which is equivalent to the state $\Omega_1$ in Eq.(\ref{equ:Omega1}). This means that the protocol can just start with the state $\Omega$ and apply the tagged model to distill the final key from the effective events using the state $\Omega_1$. The length of the final key should be
\begin{equation}\label{equ:length2}
    n_f = n_1 [1 - H (e_1^{ph})] - n_t H(E_Z)
\end{equation}
where $n_1$ is the number of effective events with state $\Omega_1$ estimated by decoy-state method, $n_t$ is the number of effective events in $Z$-windows, and $E_Z$ is the bit-flip error rate of $n_t$.
A bit-flip error occurs when Alice's bit value is different from Bob's in a $Z$-window.

{\em Note 2:Reduction of preshared information of time windows.}
\\Alice (Bob) determines a signal window by probability $p^Z_{A}$ ($p^Z_{B}$) and determines a decoy window with intensity $\mu_{Ak}$ ($\mu_{Bk}$) by probability $p^X_{Ak}$ ($p^X_{Bk}$), where $p^Z_{A}+\sum_k p^X_{Ak}=1$ ($p^Z_{B}+\sum_k p^X_{Bk}=1$).
A $Z$-window is defined when both of {\em them} determine signal windows and an $X$-window is defined when both of {\em them} determine decoy windows.
Other time windows are regarded as mismatch windows and they will be discarded.
In this way, {\em they} don't need to preshare any information of time windows, and the states in $Z$-windows and $X$-windows are $\Omega$ and $\rho_{Xk}$, respectively.

With the reductions above, the virtual protocol is equivalent to our asymmetric SNS protocol.

\section{Numerical Simulation With Asymmetric Channels}\label{sec:simulation}
Here we present the results of numerical simulation of different SNS protocols in the case that the channels are asymmetric, i.e., Alice' channel and Bob's channel are different, e.g., they have different channel losses.

The original SNS with the asymmetric channels can be modified a little to fit the asymmetric channels, which we call ``the modified SNS protocol'' in the following.
Consider the case that the original SNS protocol is applied to the asymmetric channels, when {\em they} use the same source, the intensities of the pulses interfering at the beam-splitter will differ a lot due to different channel transmittances, which will cause a high error rate in $X$-windows and therefore enhance the phase-flip error rate of single-photon, $e_1^{ph}$.
In the modified SNS protocol, {\em they} still use the same source parameters, but Charlie add an extra loss to one of the channels to make the transmittances of two channels the same.
Explicitly, if the transmittance of Alice's channel, $\eta_A$, is larger than that of Bob's channel, $\eta_B$, Charlie adds an extra loss $1-\eta_B/\eta_A$ to Alice's channel.
On the contrary, if $\eta_A<\eta_B$, Charlie adds an extra loss $1-\eta_A/\eta_B$ to Bob's channel.
Since Charlie's action doesn't affect the security, the security of the modified SNS protocol is guaranteed automatically.

\begin{figure}[!htb]
    \includegraphics[width=250pt]{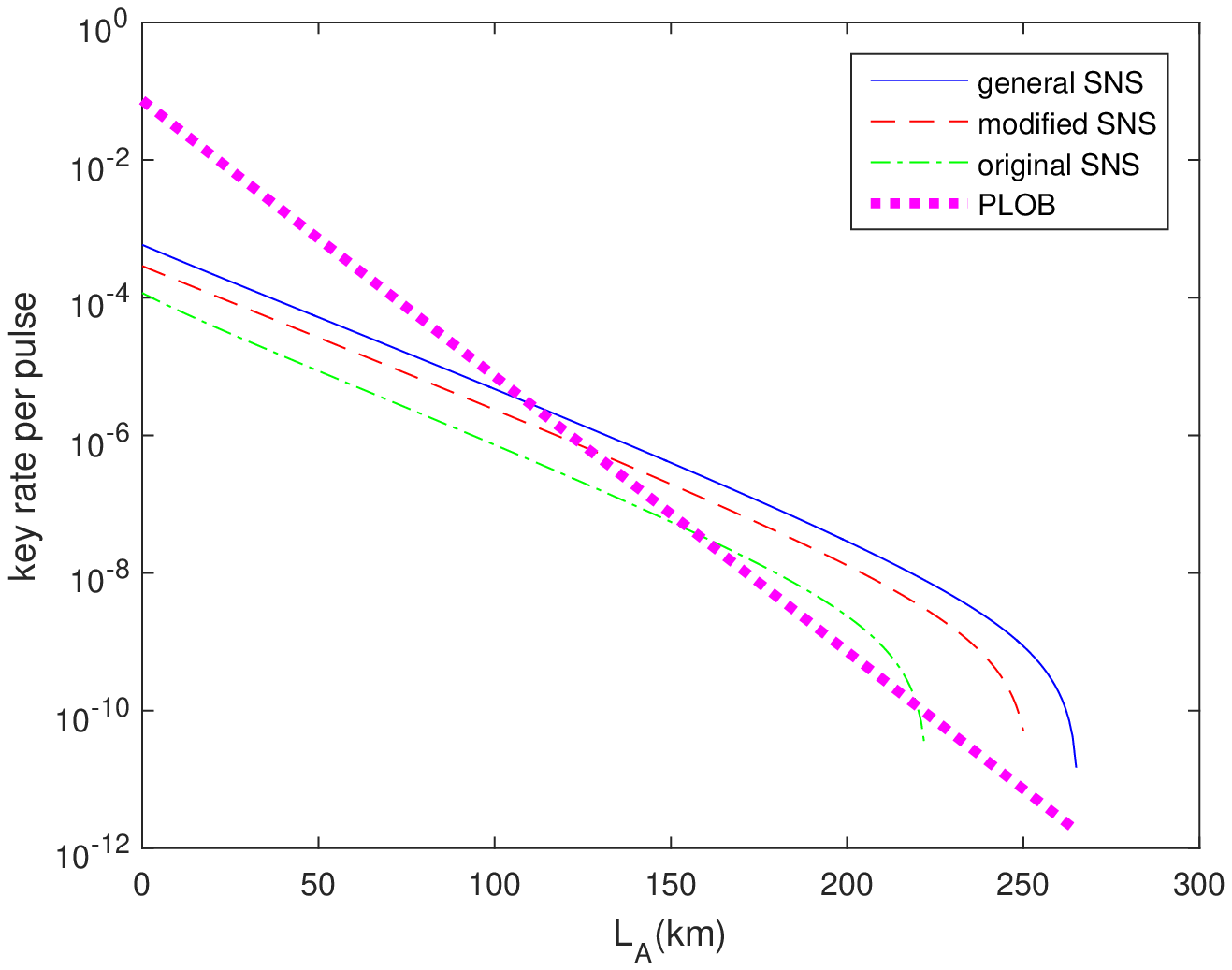}
    \caption{(Color online)The optimized key rates (per pulse pair) versus transmission distance between Alice and Charlie with three different SNS protocols. Here the difference in length between Alice's and Bob's channels is fixed at 50 km. }\label{fig:50km}
\end{figure}

\begin{figure}[!htb]
    \includegraphics[width=250pt]{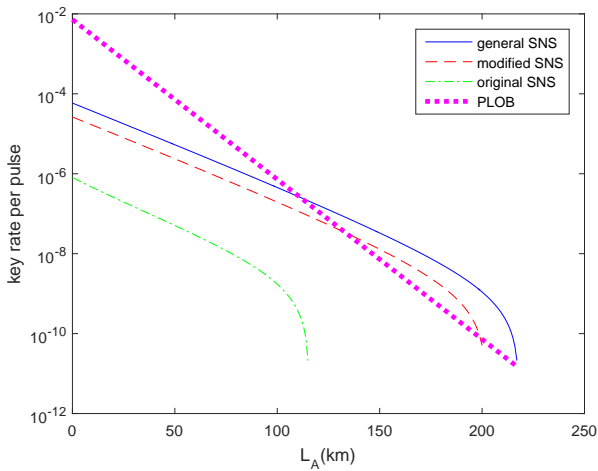}
    \caption{(Color online)The optimized key rates (per pulse pair) versus transmission distance between Alice and Charlie with three different SNS protocols. Here the difference in length between Alice's and Bob's channels is fixed at 100 km. }\label{fig:100km}
\end{figure}

We show the numerical results of the optimal key rate of the original, the modified, and the general SNS protocols with the asymmetric channels.
The effect of the finite data size has been considered in our calculation.
The device parameters used in the simulation are listed in Table~\ref{tab:parameters}.
\begin{table}
\begin{ruledtabular}
\begin{tabular}{ccccccc}
     $N_t$      &   $e_d$   &   $d$         &   $\eta_d$    &   $f_e$   &    $\xi$  &  $\alpha$ \\
    \hline
     $10^{13}$  &   5\%     &   $10^{-10}$  &   50\%        &   1.1     &   $10^{-10}$      &  0.2dB/km\\

\end{tabular}
\end{ruledtabular}
\caption{\label{tab:parameters}
    Devices' parameters used in numerical simulations.
    $N_t$ is the total number of pulse pairs;
    $e_d$ is the misalignment error in the $X$ windows;
    $d$ is the dark count rate of each detector at the UTP;
    $\eta_d$ is the detection efficiency of each detector at the UTP;
    $f_e$ is the error correction inefficiency.
    $\xi$ is the failure probability in the parameter estimation;
    $\alpha$ is the channel loss.}
\end{table}
In Figure~\ref{fig:50km} and Figure~\ref{fig:100km}, we show the optimal key rates of three protocols with the difference in length between the two channels ($L_B-L_A$) fixed at 50 km and 100 km, respectively.
In the figures, we have also compared our results with the linear bound of the repeaterless QKD.
There are excellent theoretical linear bounds for key rate of a repeaterless QKD, such as the famous TGW bound~\cite{takeoka2014fundamental} and the PLOB bound~\cite{pirandola2017fundamental}.
Also, we show some details of the optimal key rates with different SNS protocols in Table~\ref{tab:keyrate}.
\begin{table}
\begin{ruledtabular}
\begin{tabular}{c|c|c|c|c}
     $L_A$(km)  &   $L_B$(km)   &   general SNS &   modified SNS    &   original SNS \\
    \hline
        0       &   50          &   $7.21\times10^{-4}$  &   $2.89\times10^{-4}$      &   $1.17\times10^{-4}$     \\
    \hline
       150      &   200         &   $4.73\times10^{-7}$  &   $1.95\times10^{-7}$      &   $5.52\times10^{-8}$     \\
    \hline
       250      &   300         &   $1.19\times10^{-9}$  &  $5.08\times10^{-11}$      &   $0$     \\
    \hline
       0        &   100         &   $8.89\times10^{-5}$  &   $2.63\times10^{-5}$      &   $8.09\times10^{-7}$     \\
    \hline
       100      &   200         &   $6.71\times10^{-7}$  &   $1.95\times10^{-7}$      &   $1.73\times10^{-9}$     \\
    \hline
       200      &   300         &   $2.09\times10^{-9}$  &   $5.08\times10^{-11}$      &   $0$     \\
\end{tabular}
\end{ruledtabular}
\caption{\label{tab:keyrate}
    The optimal key rates with different SNS protocols. The device parameters used in the simulation are listed in Table~\ref{tab:parameters}.}
\end{table}
It's easy to find that in the asymmetric channels, the performance of our general SNS protocol is much better than that of the other two protocols, especially when the difference in length between Alice's and Bob's channels is large.

\section{Conclusion}\label{sec:conclusion}
In this paper, we propose a general SNS protocol with asymmetric source parameters and give a security proof of this protocol.
The intensities and the probabilities for sending at Alice's and Bob's sides should satisfy the condition given in Eq.(\ref{equ:protocol1}) to guarantee the security in the asymmetric case.
We present the numerical results of different SNS protocols to show that our general SNS protocol gives much high key rate than the other SNS protocols when Alice's and Bob's channels are different.
When the difference in length between Alice's and Bob's channels is 100 km, the key rate of the general SNS protocol is tens to hundreds of times higher than that of the original SNS protocol.
Our general SNS protocol can be applied directly to the SNS experiments with asymmetric channels.

If we use the method of two-way classical communication~\cite{xu2019general} on our protocol, the key rate of our general protocol can be improved further. We shall report this elsewhere.

\section*{Appendix: Formulas For Key Rate Calculation}\label{sec:formulas}
\subsection{Four-Intensity Decoy-State Method and Parameter Estimation}
In order to make our protocol easy to demonstrate in the experiment, we give the four-intensity decoy-state method for our protocol.
``Four-intensity'' means that Alice (Bob) uses four different intensities, $\mu_A^\prime$ ($\mu_B^\prime$) in signal windows and $0$, $\mu_{A1}$ ($\mu_{B1}$), $\mu_{A2}$ ($\mu_{B2}$) in decoy windows.
All measurement results in effective $X$-windows are used to estimate the bound of $n_1$.
Only measurement results in effective $\tilde{X}$-windows when Alice uses the intensity $\mu_{A1}$ and Bob uses the intensity $\mu_{B1}$ are used to estimate the bound of $e_1^{ph}$.

The formula for the length of the final key in Eq.(\ref{equ:length}) can be written in the form of key rate per time window:
\begin{equation}\label{equ:keyrate}
\begin{split}
    R = p_A^Z p_B^Z \{ &[\epsilon_A(1-\epsilon_B) \mu_A^\prime e^{-\mu_A^\prime} + \epsilon_B(1-\epsilon_A) \mu_B^\prime e^{-\mu_B^\prime}]\\
     & s_1^Z [1 - H(e_1^{ph})] - f S_Z H(E_Z) \}
\end{split}
\end{equation}
where $s_1^Z$ is the counting rate in $Z_1$-windows, and $S_Z$ is the counting rate of $Z$-windows. If there are $m$ effective windows in a set $\zeta$ of $n$ time windows, the counting rate of $\zeta$ is defined as $S_\zeta=m/n$.

So we need to estimate the bound of $\mean{s_1^Z}$ and $\mean{e_1^{ph}}$ by the four-intensity decoy-state method. Here $\mean{\cdot}$ stands for the expected value of a variable. Similarly to the methods in Ref.\cite{yu2019sending}, the lower bound of $\mean{s_1^Z}$ is given by:
\begin{equation}\label{equ:s1Z}
    \mean{s_1^Z} \ge \underline{\mean{s_1^Z}} = \frac{\mu_{A1}}{\mu_{A1}+\mu_{B1}} \underline{\mean{s_{10}^Z}} + \frac{\mu_{B1}}{\mu_{A1}+\mu_{B1}} \underline{\mean{s_{01}^Z}}
\end{equation}
where $\underline{\mean{s_{10}^Z}}$ is lower bound of the expected value of the counting rate of the state $\oprod{10}{10}$ with
\begin{equation}\label{equ:s10}
    \underline{\mean{s_{10}^Z}} \!=\! \frac{\mu_{A2}^2 e^{\mu_{A1}} \mean{S_{\mu_{A1}0}} \!-\! \mu_{A1}^2 e^{\mu_{A2}} \mean{S_{\mu_{A2}0}} \!-\! (\mu_{A2}^2 \!-\! \mu_{A1}^2) \mean{S_{00}}}{\mu_{A1} \mu_{A2} (\mu_{A2} - \mu_{A1})}
\end{equation}
and $\underline{\mean{s_{01}^Z}}$ is lower bound of the expected value of the counting rate of the state $\oprod{01}{01}$ with
\begin{equation}\label{equ:s01}
    \underline{\mean{s_{01}^Z}} \!=\! \frac{\mu_{B2}^2 e^{\mu_{B1}} \mean{S_{0\mu_{B1}}} \!-\! \mu_{B1}^2 e^{\mu_{B2}} \mean{S_{0\mu_{B2}}} \!-\! (\mu_{B2}^2 \!-\! \mu_{B1}^2) \mean{S_{00}}}{\mu_{B1} \mu_{B2} (\mu_{B2} - \mu_{B1})}.
\end{equation}
Here $\mean{S_{\alpha\beta}}$ is the expected value of the counting rate of the time windows when Alice and Bob send the decoy pulses with intensities $\alpha$ and $\beta$, respectively.
If the data size is infinite, these expected values are exactly the values observed in the experiments.
If the data size is finite, we should use the Chernoff Bound introduced in the next section to estimate the bound of the expected values from the observed values.
The upper bound of $\mean{e_1^{ph}}$ is given by:
\begin{equation}\label{equ:e1ph}
    \mean{e_1^{ph}} \le \overline{\mean{e_1^{ph}}} = \frac{\mean{T_{\Delta}} - e^{-(\mu_{A1}+\mu_{B1})} \mean{S_{00}} /2}{e^{-(\mu_{A1}+\mu_{B1})} (\mu_{A1}+\mu_{B1}) \underline{\mean{s_1^Z}}}
\end{equation}
where $\mean{T_{\Delta}}$ is the expected value of the error counting rate of $\tilde{X}$-windows, when Alice and Bob send the decoy pulses with intensities $\mu_{A1}$ and $\mu_{B1}$, respectively.
If there are $m$ error windows in a set $\zeta$ of $n$ time windows, the error counting rate of $\zeta$ is defined as $T_\zeta=m/n$.

\subsection{Chernoff Bound}
In the asymptotic case where the data size is infinite, the observed values are the same as the expected values.
But in the nonasymptotic case where the data size is finite, the observed values are different from the expected values.
So we need the Chernoff bound~\cite{chernoff1952measure} to estimate the range of expected values from the observed values and use the worst case to ensure that the final key is secure.

Let $X_1,X_2,\dots,X_n$ be $n$ random variables whose observed values are either 0 or 1, $X$ be their sum $X=\sum_i X_i$, and $\phi$ be the expected value of $X$. We have the lower and the upper bound of $\phi$:
\begin{equation}\label{equ:chornoffL}
    \phi^L (X) = \frac{X}{1+\delta_1(X)}
\end{equation}
\begin{equation}\label{equ:chornoffU}
    \phi^U (X) = \frac{X}{1-\delta_2(X)}
\end{equation}
where $\delta_1(X)$ and $\delta_2(X)$ are the solutions of the following equations:
\begin{equation}\label{equ:delta1}
    \left(\frac{e^{\delta_1}}{(1+\delta_1)^{1+\delta_1}}\right)^{\frac{X}{1+\delta_1}} = \frac{\xi}{2}
\end{equation}
\begin{equation}\label{equ:delta2}
    \left(\frac{e^{-\delta_2}}{(1-\delta_2)^{1-\delta_2}}\right)^{\frac{X}{1-\delta_2}} = \frac{\xi}{2}
\end{equation}
where $\xi$ is the failure probability. With the above equations, we have
\begin{equation}\label{equ:bound1}
    N_{\alpha\beta} \underline{\mean{S_{\alpha\beta}}} = \phi^L(N_{\alpha\beta} S_{\alpha\beta}),N_{\alpha\beta} \overline{\mean{S_{\alpha\beta}}} = \phi^U(N_{\alpha\beta} S_{\alpha\beta}).
\end{equation}
Here $S_{\alpha\beta}$ is the observed value of the counting rate.

Then in Eq.(\ref{equ:keyrate}) we need the real values of $s_1^Z$ and $e_1^{ph}$ in a specific experiment. So Eqs.(\ref{equ:chornoffL})-(\ref{equ:bound1}) can be written in another form to estimate the upper and the lower bound of real values from expected values:
\begin{equation}\label{equ:chornoffU'}
    X^U(\phi) = [1+\delta_1^\prime(\phi)]\phi
\end{equation}
\begin{equation}\label{equ:chornoffL'}
    X^L(\phi) = [1-\delta_2^\prime(\phi)]\phi
\end{equation}
where $\delta_1^\prime(\phi)$ and $\delta_2^\prime(\phi)$ are the solutions of the following equations:
\begin{equation}\label{equ:delta1'}
    \left(\frac{e^{\delta_1^\prime}}{(1+\delta_1^\prime)^{1+\delta_1^\prime}} \right)^{\phi} = \frac{\xi}{2}
\end{equation}
\begin{equation}\label{equ:delta2'}
    \left(\frac{e^{-\delta_2^\prime}}{(1-\delta_2^\prime)^{1-\delta_2^\prime}} \right)^{\phi} = \frac{\xi}{2}
\end{equation}
With the above equations, we have
\begin{equation}\label{equ:bound2}
\begin{split}
    N_1^Z s_1^Z &\ge X^L(N_1^Z \underline{\mean{s_1^Z}}),\\
    N_1^Z \underline{\mean{s_1^Z}} e_1^{ph} &\le X^U(N_1^Z \underline{\mean{s_1^Z}} \overline{\mean{e_1^{ph}}}).
\end{split}
\end{equation}
where $N_1^Z=N_t p_A^Z p_B^Z [\epsilon_A(1-\epsilon_B) \mu_A^\prime e^{-\mu_A^\prime} + \epsilon_B(1-\epsilon_A) \mu_B^\prime e^{-\mu_B^\prime}]$ is the number of single-photon state in the $Z$ windows when one and only one of {\em them} decides to send and $N_t$ is the total number of time windows.

\subsection{Finite Key Size Effect}
Similarly to the analysis of the effect of the finite key size in ref.~\cite{jiang2019unconditional}, we give the key rate formula with the effect of the finite key size of our general SNS protocol in the universally composable framework~\cite{muller2009composability}.

If the length of the final key satisfies
\begin{equation}\label{equ:finitekey}
\begin{split}
    N_f = n_1 [1 - H(e_1^{ph})] - f n_t H(E_Z) \\
     - \log_2 \frac{2}{\varepsilon_{\text{cor}}} - 2\log_2 \frac{1}{\sqrt2 \varepsilon_{\text{PA}} \hat\varepsilon},
\end{split}
\end{equation}
the protocol is $\varepsilon_{\text{sec}}$-secret with $\varepsilon_{\text{sec}}=2\hat\varepsilon+4\bar\varepsilon+\varepsilon_{\text{PA}}+\varepsilon_{n_1}$ , and the total security coefficient of the protocol is $\varepsilon_{\text{tot}}=\varepsilon_{\text{cor}}+\varepsilon_{\text{sec}}$.
Here $\varepsilon_{\text{cor}}$ is the probability that the error correction fails, $\bar\varepsilon$ is the probability that the real value of $e_1^{ph}$ isn't in the range that we estimate, $\varepsilon_{\text{PA}}$ is the failure probability of the privacy amplification, and $\varepsilon_{n_1}$ is the probability that the real value of $n_1$ isn't in the range that we estimate.
According to Eqs.(\ref{equ:s1Z})-(\ref{equ:bound2}), we have $\bar\varepsilon=3\xi$ and $\varepsilon_{n_1}=6\xi$.
If we set $\varepsilon_{\text{cor}}=\hat\varepsilon=\varepsilon_{\text{PA}}=\xi$ in our numerical simulation, the total security coefficient of our protocol is $\varepsilon_{\text{tot}}=22\xi=2.2\times10^{-9}$.

Also, Eq.(\ref{equ:finitekey}) can be written in the form of key rate per time window with some source parameters:
\begin{equation}\label{equ:finitekeyrate}
\begin{split}
    R = p_A^Z p_B^Z \{ &[\epsilon_A(1-\epsilon_B) \mu_A^\prime e^{-\mu_A^\prime} + \epsilon_B(1-\epsilon_A) \mu_B^\prime e^{-\mu_B^\prime}]\\
    &\cdot s_1^Z [1 - H(e_1^{ph})] - f S_Z H(E_Z) \} \\
      &- \frac{1}{N_t} (\log_2 \frac{2}{\varepsilon_{\text{cor}}} + 2\log_2 \frac{1}{\sqrt2 \varepsilon_{\text{PA}} \hat\varepsilon}).
\end{split}
\end{equation}
\newline

\bibliography{refs}

%
%

\end{document}